\documentclass[fleqn,usenatbib]{mnras}

\DeclareRobustCommand{\VAN}[3]{#2}
\let\VANthebibliography\thebibliography
\def\thebibliography{\DeclareRobustCommand{\VAN}[3]{##3}\VANthebibliography}

\usepackage{graphicx}	% Including figure files
\usepackage{amsmath}	% Advanced maths commands
\usepackage{amssymb}	% Extra maths symbols

\usepackage{newtxtext,newtxmath}
\usepackage[T1]{fontenc}

\DeclareMathOperator{\sech}{sech}
\newcommand{\vlos}{v_\mathrm{los}}
\newcommand{\bfx}{\bmath{x}}
\newcommand{\bfy}{\bmath{y}}
\newcommand{\D}{\mathcal{D}}

\usepackage{xcolor}
\newcommand{\hl}[1]{\textcolor{magenta}{#1}}
\renewcommand{\hl}[1]{#1}

\title[The missing radial velocities of \textit{Gaia}]{The missing radial velocities of \textit{Gaia}: blind predictions for DR3}

\author[A. P. Naik et al.]{
Aneesh P. Naik$^{1}$\thanks{E-mail: aneesh.naik@nottingham.ac.uk}
and Axel Widmark$^{2}$
\\
$^{1}$School of Physics \& Astronomy, University of Nottingham, University Park, Nottingham NG7 2RD, UK\\
$^{2}$Dark Cosmology Centre, Niels Bohr Institute, University of Copenhagen, Jagtvej 128, 2200 Copenhagen N, Denmark
}

\date{Accepted XXX. Received YYY; in original form ZZZ}

\pubyear{2022}

\begin{document}
\label{firstpage}
\pagerange{\pageref{firstpage}--\pageref{lastpage}}
\maketitle

%%%%%%%%%%%%%%%%%%%%%%%%%%%%%%%%%%%%%%%
% abstract & keywords

\begin{abstract}
While \textit{Gaia} has observed the phase space coordinates of over a billion stars in the Galaxy, in the overwhelming majority of cases it has only obtained five of the six coordinates, the missing dimension being the radial (line-of-sight) velocity. Using a realistic mock dataset, we show that Bayesian neural networks are highly capable of `learning' these radial velocities as a function of the other five coordinates, and thus filling in the gaps. For a given star, the network outputs are not merely point predictions, but full posterior distributions encompassing the intrinsic scatter of the stellar phase space distribution, the observational uncertainties on the network inputs, and any `epistemic' uncertainty stemming from our ignorance about the stellar phase space distribution. Applying this technique to the real \textit{Gaia} data, we generate and publish a catalogue of posteriors \hl{(median width: 25~km/s)} for the radial velocities of 16 million \textit{Gaia} DR2/EDR3 stars in the magnitude range $6<G<14.5$. Many of these gaps will be filled in very soon by \textit{Gaia} DR3, which will serve to test our blind predictions. Thus, the primary use of our published catalogue will be to validate our method, justifying its future use in generating an updated catalogue of posteriors for radial velocities missing from \textit{Gaia} DR3.
\end{abstract}

\begin{keywords}
Galaxy: kinematics and dynamics -- catalogues -- techniques: radial velocities -- methods: statistical
\end{keywords}

%%%%%%%%%%%%%%%%%%%%%%%%%%%%%%%%%%%%%%%
% intro
\section{Introduction}

In the past few years, the observations made by the \textit{Gaia} satellite \citep{Gaia2016mission} have brought about a golden age of studying and understanding the composition, history, and dynamics of our Galaxy. Often publicised as the `billion-star surveyor', \textit{Gaia} has observed the positions and motions of well over a billion stars, most of them within a few kiloparsecs of the Sun. These data have been published in 2.5 data releases: Data Release 1 in 2016 \citep[DR1;][]{Gaia2016DR1}, Data Release 2 in 2018 \citep[DR2;][]{Gaia2018DR2}, and Early Data Release 3 in late 2020 \citep[EDR3;][]{Gaia2021EDR3}, with the full Data Release 3 (DR3) due to be published in the week following the submission of this article (June 2022).

Despite the richness and abundance of the \textit{Gaia} data, analyses of Galactic dynamics are often stymied by missing information. For the overwhelming majority of stars, \textit{Gaia} only provides five of the six phase space coordinates: two sky coordinates, a parallax, and two proper motions. As of EDR3, the sixth coordinate, velocity along the line of sight or `radial velocity' $\vlos$ is only available for around seven million stars \citep[with most measurements carried over from DR2;][]{Seabroke2021}, which is only a small fraction of the wider \textit{Gaia} sample. This is due to a number of selection effects, chiefly the Radial Velocity Spectrometer (RVS) aboard the \textit{Gaia} satellite having a brighter magnitude limit than the astrometric and photometric instruments \citep{Cropper2018, Katz2019}.

Under these circumstances, if one wishes to study the dynamics of the \textit{Gaia} stars, there are several options. First, one could limit the analysis to stars with complete phase space information, and model the associated selection effects if necessary. However, this option is limiting and does not utilise the full statistical power of the vast dataset. Second, one could use the 5D stars to study the spatial part of the phase space distribution and the 6D stars to study the velocity part \citep[e.g.,][]{Salomon2020, Naik2022}. However, the 5D stars being discarded for the velocity analysis do have two of their three velocity measurements available, so useful information is being lost. 

The final option is to use all stars, adopting informed guesses or marginalising over probability distributions for the radial velocities of the 5D stars \citep[e.g.,][]{Widmark2021}. In that vein, it would be useful to have estimates (or better still, predictive probability distributions) for the 5D \textit{Gaia} stars, but no such estimates have been published to date. Producing such estimates is not a straightforward undertaking: to understand how the radial velocity of a given star depends on its other five phase space coordinates, one would need to first understand the overall stellar phase space distribution function (DF). The Milky Way's stars constitute a rich, heterogeneous population that is both spatially and kinematically highly complex, and a myriad of asymmetries and substructures have been observed in various projections of phase space \citep[e.g.,][]{Schoenrich2018, Gaia2018disc, Antoja2018, Bennett2019, Salomon2020}. Given this complexity, an analytic DF that truly captures the full six-dimensional richness of the distribution and motions of the Galaxy's stars would be very difficult to write down. On the other hand, the sheer size of the dataset available from \textit{Gaia} ought to give an appreciable degree of constraining power if one instead adopts a more data-driven, machine learning approach to the problem \citep{Naik2022, Green2022, Buckley2022}.

Using just such an approach, \citet{Dropulic2021} took the first steps towards estimating the radial velocities of the 5D \textit{Gaia} stars. Under their formulation, an artificial neural network (ANN) `learns' the dependence of both $\vlos$ and an uncertainty $\sigma_{\vlos}$ on the other five phase space coordinates. Given predictions for these quantities, one can subsequently sample specific radial velocities for each 5D star from one-dimensional Gaussian distributions of mean $\vlos$ and width $\sigma_{\vlos}$. In their article, \citet{Dropulic2021} demonstrated that this technique is capable of faithfully reproducing the velocity distributions of stars drawn from the \textit{Gaia} DR2-like mock catalogue of \citet{Rybizki2018}. \hl{In an impressive follow-up article, \citet{Dropulic2022} applied their technique to the real \textit{Gaia} data and used the resulting $\vlos$ predictions to study the accreted population from the \textit{Gaia}-Sausage/Enceladus merger \citep{Belokurov2018, Helmi2018}; by estimating these missing radial velocities, they were able to increase the number of GSE candidates in the solar neighbourhood by a factor of twenty, illustrating the benefits of filling in the sixth dimension.}

In this work, we also use a technique from deep learning to predict the radial velocities of the 5D \textit{Gaia} stars. We extend the work of \citet{Dropulic2021, Dropulic2022} in two key respects. First, we use Bayesian neural networks \citep[BNNs;][]{Titterington2004, Goan2020, Jospin2020} instead of classical, deterministic ANNs. Even when used to fit non-deterministic probability distributions as in \citet{Dropulic2021, Dropulic2022}, a significant limitation of deterministic ANNs is the danger of over-fitting the data in areas where the data are scarce (as could be the case here in some under-sampled regions of phase space), although countermeasures exist, such as regularisation. With BNNs, the model parameters (i.e. the internal network parameters) are promoted from fixed parameters to random variables. Given some data and a prior distribution for the model parameters, one trains a BNN by learning the posterior distribution. When it comes to subsequently predicting new values given some new inputs, BNNs do not merely generate point predictions but samples from full posterior predictive probability distributions marginalising over the posteriors of the model parameters. These predictive distributions represent, in a fully Bayesian spirit, our current state of knowledge given the data as well as our prior knowledge (or ignorance) about the model parameters. By this very nature, BNNs are significantly less prone to over-fitting the data than deterministic ANNs \citep{Mitros2019, Ovadia2019, Kristiadi2020, Jospin2020}. The second advance our methodology makes beyond that of \citet{Dropulic2021} is that our posteriors are not constrained to being Gaussian distributions, and are instead highly flexible, allowing skewed, pinched, or even multi-modal posteriors.

%The third and final way in which we extend the work of \citet{Dropulic2021} is that we go beyond mock data and apply our technique to the real \textit{Gaia} data, 

We construct a catalogue of predictions for all EDR3 stars with $6 < G < 14.5$ that have accompanying photo-astrometric distance estimates from the \texttt{StarHorse} code \citep{Anders2022}. We submit this article and publish the accompanying catalogue less than a week prior to the publication of \textit{Gaia} DR3. DR3 will publish radial velocities for stars down to fainter magnitudes than DR2/EDR3, and will thus overlap significantly with our catalogue and test our predictions, thus justifying the use of our method in the near future to generate probability distributions for the line-of-sight velocities of the 5D stars in DR3.

This paper is structured as follows. In the following section (Sec.~\ref{S:BNN}), we give a brief overview of BNNs and the technical details of our implementation. Subsequently, we show the results of training BNNs on a mock EDR3-like dataset (Sec.~\ref{S:MockEDR3}) and the real Gaia EDR3 (Sec.~\ref{S:EDR3}), before providing some concluding remarks in Sec.~\ref{S:Conclusions}.

%%%%%%%%%%%%%%%%%%%%%%%%%%%%%%%%%%%%%%%
% BNNs
\section{Bayesian Neural Networks}
\label{S:BNN}

\begin{table}
\centering
\caption{Various symbols and quantities used in Sec.~\ref{S:BNN}.}
\begin{tabular}{l | l}
\hline
\textbf{BNN architecture}  & \\
\hline
$N_n$                 & Number of units in network layer $n$\\
$\mathbfss{W}^{(n)}$  & Matrix of weights of layer $n$ (shape $N_n \times N_{n-1}$)\\
$\bmath{b}^{(n)}$     & Vector of biases of layer $n$ (length $N_n$)\\
$s^{(n)}$             & Activation function of layer $n$ ($\mathbb{R}^{N_n} \rightarrow \mathbb{R}^{N_n}$)\\
${\bf z}^{(n)}$       & Output vector of layer $n$ (length $N_n$; see Eq.~\ref{E:NNLayer})\\
$f_\theta(\bfx)$      & Final network output vector\\
\hline
\textbf{Data}  & \\
\hline
$\mathcal{D}$    & Set of training data, partitioned into $\D_x, \D_y$\\
$\mathcal{D}_x$  & Independent variable columns (`features') of $\D$\\
$\mathcal{D}_y$  & Dependent variable columns (`labels') of $\D$\\
$N$              & Number of rows in training data\\
$\bfx$           & Single row of $\D_x$ (i.e. network input vector)\\
$\bfy$           & Single row of $\D_y$\\
$\bfx'$          & Network inputs for data unseen in training\\
$\bfy'$          & Labels for data unseen in training\\
\hline
\textbf{Sampling} & \\
\hline
$\mu(X)$     & Mean of random variable $X$\\
$\sigma(X)$  & Standard deviation of random variable $X$\\
$N_s$        & Number of samples taken\\
\hline
\textbf{Parameters} & \\
\hline
$\theta$    & Network weights and biases $\{\mathbfss{W}, \bmath{b}\}$\\
$\psi$      & $\theta$ PDF parameters $\{\mu(\theta), \sigma(\theta)\}$\\
$\hat\psi$  & Optimal $\psi$ such that $p(\theta|\psi) \approx p(\theta | \D)$\\
\hline
\textbf{Indices} & \\
\hline
$i$  & Labels individual data points $\bfx_i, \bfy_i$ of $\D$\\
$j$  & Labels posterior samples \\
\end{tabular}
\label{T:Symbols}
\end{table}

Our ultimate goal is to derive, for each 5D \textit{Gaia} star, a posterior probability distribution for its radial velocity given its five other phase space coordinates. This posterior should reflect the intrinsic scatter of the stellar phase space distribution, the observational uncertainties on the inputs, and also any epistemic uncertainty stemming from a lack of data. We achieve this using Bayesian neural network (BNNs).

In this section we give a brief overview of some BNN fundamentals and the details of our implementation. For reference, we provide a list of symbols and indices used throughout this section in Table~\ref{T:Symbols}. For a more comprehensive and pedagogical introduction to BNNs, we refer the reader to \citet{Jospin2020}.

\subsection{Premise}
\label{S:BNN:Premise}

Let us first understand the operation of a classical artificial neural network (ANN) which produces deterministic point predictions for given inputs. Ultimately, an ANN represents some arbitrary, highly complex model function $f_\theta(\bfx)$, where $\theta$ is a set of parameters and the inputs $\bfx$ and outputs $f_\theta(\bfx)$ are vectors (not necessarily of the same length). In the textbook example of a feedforward ANN, the network is commonly visualised as a sequence of layers: an input layer, a specified number of hidden layers (each comprising a specified number of `units'), and an output layer. When an input vector $\bfx$ (length $N_0$) is fed to the ANN, it is transformed at the first hidden layer (comprising $N_1$ units) into a vector $\bmath{z}^{(1)}$ (length $N_1$) via
\begin{equation}
\label{E:NNLayer}
    \bmath{z}^{(1)} = s^{(1)}(\mathbfss{W}^{(1)}\bfx + \bmath{b}^{(1)}),
\end{equation}
where the components of the matrix $\mathbfss{W}^{(1)}$ (shape $N_1 \times N_0$) and vector $\bmath{b}^{(1)}$ (length $N_1$) are known as the `weights' and `biases' of this layer, and $s^{(1)}$ is some arbitrary non-linear function, which in general can be chosen differently for each layer. The vector $\bmath{z}^{(1)}$ then continues to the next hidden layer (comprising $N_2$ units) where it is subject to a similar transformation into $\bmath{z}^{(2)}$ (length $N_2$), but now with a new matrix of weights (shape $N_2 \times N_1$) and vector of biases (length $N_2$). This process continues through the hidden layers until a final transformation into vector $f_\theta(\bfx)$ at the output layer. In summary, the function $f_\theta$ comprises a nested series of non-linear functions applied to linear combinations of inputs. The parameter set $\theta=\{\mathbfss{W}^{(1)}, \bmath{b}^{(1)}, \mathbfss{W}^{(2)}, \bmath{b}^{(2)}, ...\}$ then comprises all of the weights and biases across the layers of the network. Given such an ANN, one typically `trains' the network with some training data (i.e. a series of `features' $\bfx$ and `labels' $\bfy$) by optimising the free parameters $\theta$ to minimise some given `loss function' which quantifies the difference between the labels $\bfy$ and the model predictions $f_\theta(\bfx)$.

BNNs are fundamentally similar to ANNs in operation, but they introduce stochasticity to the process, such that on each forward pass for a given input, the network will generate a different output. By generating a number of such outputs, one constructs not just a point prediction but a series of predictions which can be understood as samples from a probability distribution for $\bfy$. If a BNN has been trained on data $\D$ (comprising a series of $\bfx, \bfy$ feature/label pairs), then given a new input $\bfx'$, the sampled probability distribution for the output $\bfy'$ can be understood as the posterior predictive distribution $p(\bfy' | \bfx', \D)$. 

Typically, this stochasticity is introduced to the neural network by promoting the weights $\mathbfss{W}$ and biases $\bmath{b}$ of the network fixed parameters to random variables with some choice of probability distribution, and so the network outputs vary on each forward pass because the weights and biases are resampled from their probability distributions each time. The governing parameters $\psi$ of these probability distributions are then the free parameters to be optimised, rather than the weights and biases directly. We follow most BNN implementations in using 1D Gaussian distributions for each weight and bias, so that $\psi=\{\mu({W^{(1)}_{11}}), \sigma(W^{(1)}_{11}), \mu({b^{(1)}_{1}}), \sigma(b^{(1)}_{1}), \mu({W^{(1)}_{12}}), ...\}$, where $\mu(X)$ and $\sigma(X)$ indicate the mean and standard deviation of variable $X$. This 1D treatment assumes zero covariances between difference weights and biases. This is primarily for computational ease: using more general multivariate Gaussians would prove extremely resource intensive given the large number of weights and biases in a typical neural network.

\subsection{Optimisation}
\label{S:BNN:Optimisation}

Different procedures for optimising BNNs exist, with perhaps the most popular being the `Bayes-by-backprop' \citep{Blundell2015} algorithm. However, applied to toy problems and to the present radial velocity problem, we found that this technique consistently led to BNN predictions that were either over-confident or posteriors that were broad enough to encompass the whole dataset, depending on network hyper-parameters. For this reason, we have instead formulated a novel optimisation procedure which we summarise here. We describe the procedure of optimising $\psi$ given a training set $\D$ comprising `features' $\D_x = \{\bfx\}$ (i.e. the network inputs) and `labels' $\D_y = \{\bfy\}$ (i.e. the `truths' to be compared to the network outputs $f_\theta(\bfx)$). One begins with some initial value for the PDF parameters $\psi$, then repeats the following steps as required:
\begin{enumerate}
    \item Using the current PDF parameters $\psi$, sample a number $N_s$ of network parameter sets $\{\theta_j\}, j=1,2,...,N_s$, sampling each weight and bias in $\theta$ from its 1D Gaussian. Note that this is implicitly assuming an infinite flat prior for $\theta$. To adopt a different prior $p(\theta)$, sample instead from $p(\theta) \times \mathcal{N}\left(\mu(\theta), \sigma(\theta)\right)$.
    \item For each data point $\bfx_i$ in the training set, perform forward passes through the network for all sampled sets $\theta_j$, thus obtaining $N_s$ predictions per data point: $\{f_{\theta_j}(\hat\bfx_i)\}$.
    \item For each data point, convert this set of predictions into a smooth probability distribution $p(\bfy | \bfx_i, \psi)$. We do this with kernel density estimation using a logistic ($\sech^2$) kernel, i.e.
    \begin{equation}
    \label{E:KDE}
        p(\bfy | \bfx_i, \psi) = \frac{1}{N_s} \sum_j^{N_s}\frac{1}{4h_i} \mathrm{sech}^2\left(\frac{|\bfy - f_{\theta_j}(\bfx_i)|}{2 h_i}\right).
    \end{equation}
    Here, $h_i$ is the kernel bandwidth, which is set differently for each data point, according to
    \begin{equation}
    \label{E:Bandwidth}
        h_i = 0.6 \sigma_i N_s^{-1/5},
    \end{equation}
    where $\sigma_i$ is the standard deviation of the BNN predictions $\{f_{\theta_j}(\hat\bfx_i)\}$ for data point $i$. Eq.~\ref{E:Bandwidth} is similar to the `rule-of-thumb' for optimal bandwidth selection recommended by \citet{Silverman1986}, except that we have reduced the numerical factor 0.9 to 0.6: we found that using 0.6 with logistic kernels gives a visually similar degree of smoothing to using 0.9 with Gaussian kernels. 
    \item Given the probability distribution, evaluate the probability of observing the truth $\bfy_i$, $p(\bfy=\bfy_i | \bfx_i, \psi)$.
    \item Multiply $p(\bfy=\bfy_i | \hat\bfx, \psi)$ over all data points to give the overall probability $p(\D_y | \D_x, \psi)$.
    \item Calculate the `loss', given by the average negative log probability,
    \begin{equation}
    \label{E:Loss}
        \text{loss} = -\ln p(\D_y | \D_x, \psi)/N,
    \end{equation}
    where $N$ is the size of the dataset $\D$.
    \item Given the calculated loss, update $\psi$, e.g. via a gradient descent step.
\end{enumerate}

In Bayesian terms, the process described above can be understood as finding the optimal $\hat\psi$ such that the probablity distribution of $\theta$ (i.e. the joint probability distribution over all the weights and biases of the network) given $\psi=\hat\psi$, $p(\theta | \psi=\hat\psi)$, closely resembles the posterior probability distribution of $\theta$ given the training data $p(\theta | \D)$. Techniques of this kind are known as `variational inference'. For a review of this class of techniques, see \citet{Blei2017}.

Having thus trained the BNN, given a new input $\bfx'$, one can generate a series of samples from the posterior predictive distribution $p(\bfy' | \bfx', \D)$ by simply repeating steps (i) and (ii) of the procedure above, adopting $\psi=\hat\psi$. If desired, these samples can be further converted into a smooth probability distribution via step (iii).

\subsection{Implementation}
\label{S:BNN:Implementation}

We have thus far discussed BNNs and their optimisation in general terms. Regarding the present work in particular, we implement a feedforward BNN in \textsc{pytorch} \citep{Paszke2019}. We use the same BNN architecture for both the mock dataset (Sec.~\ref{S:MockEDR3}) and the real \textit{Gaia} data (Sec.~\ref{S:EDR3}): 8 hidden layers, each with 64 units with Rectified Linear Unit (ReLU) activation functions \citep{Nair2010}. We find that increasing these hyper-parameters gives no appreciable improvement in accuracy or precision (i.e. posterior widths). The BNN has five inputs: Galactocentric X/Y/Z; proper motions in right ascension and declination $\mu_\alpha$/$\mu_\delta$, and one output: $\vlos$. We experimented with additional inputs, in particular we separately added absolute and apparent magnitudes, but found no improvement in the results. Our conversion to Cartesian positions\footnote{To perform this conversion, we assume the distance from the Sun to the Galactic centre is 8.122~kpc \citep{GRAVITY2018} and the height of the Sun above the Galactic mid-plane is 20.8~pc \citep{Bennett2019}. For the sky coordinates (right ascension, declination) of the Galactic centre we use 266.4051\degr, -28.9362\degr~\citep{Reid2004}, and for the Galactic `roll' angle we use a value of 58.5986\degr, derived from the IAU coordinates of the Galactic north pole \citep{deBlaauw1960}.} from the right ascensions and declinations of the raw data is to avoid the angular discontinuity of a spherical polar coordinate system at zero azimuth: it would not be obvious to the BNN that stars with right ascensions of 0.1\degr and 359.9\degr are proximate on the sky. For both the mock and real datasets, we shift and rescale the data prior to training to give means of approximately zero and standard deviations of approximately one along each of the six dimensions. 

On each forward pass during the training, we draw $N_s=250$ samples for $\theta$ (step (i) in Sec.~\ref{S:BNN:Implementation}), which are then used to generate the $N_s$ predictions for each data point $\bfx_i$ (step (ii)). To minimise the loss, we use the stochastic gradient descent algorithm \textsc{adam} \citep{Kingma2015}. For both the mock data and the real \textit{Gaia} data, performing forward passes on the whole dataset at once proves too memory intensive, so we instead sub-divide the dataset into batches of size 6000. A single training `epoch' then consists of looping over the batches, evaluating the loss function (Eq.~\ref{E:Loss}) on each individual batch and updating the training parameters $\psi$ accordingly. The \textsc{adam} algorithm requires a `learning rate', essentially a gradient descent step size. We start with learning rate of $10^{-2}$, and reduce the learning rate by a factor of 2 once 10 training epochs pass without an appreciable reduction in the loss (averaged across all the batch losses calculated in a given epoch). The training is then truncated when the learning rate crosses a threshold value of $10^{-5}$, or 500 epochs have passed, whichever occurs first. In practice it is always the first criterion that is met first.

Each star has quoted observational uncertainties on its position, proper motions, and (where present) its radial velocity. To account for these uncertainties, on each training epoch we resample the positions and velocities of each star in the training set from its error distribution. Because of technical differences between the mock dataset and the real \textit{Gaia} data, the precise manner in which we do this sampling differs between the two applications, and so we will discuss this sampling in more detail in Secs.~\ref{S:MockEDR3} and \ref{S:EDR3} respectively. \hl{This procedure of generating realisations from a star's error distribution is a way to inflate our prediction widths in a manner that accounts for observational uncertainties. However, this procedure might lead to over-estimated uncertainties on the resulting predictions, as the resampling is in effect re-perturbing each measurement further from its true value after it has already been perturbed once by the measurement process. A more sophisticated technique for dealing with observational uncertainties \citep[such as `extreme deconvolution';][]{Bovy2011} could in principle lead to more precise predictions.} Furthermore, under our treatment, each realisation is given the same weight, although in principle different realisations might have very different posterior probabilities; this simplification could hinder `Bayesian shrinkage' to some degree. In other words, our model cannot account for the fact that an object with very large observational uncertainties could in principle still have its positions and motions constrained by the phase-space distribution of the overall stellar population. However, these effect is negligible as long as observational uncertainties are small, which is why in the case of the real data we train our model with some modest cuts in data quality.

\begin{figure*}
    \centering
    \includegraphics{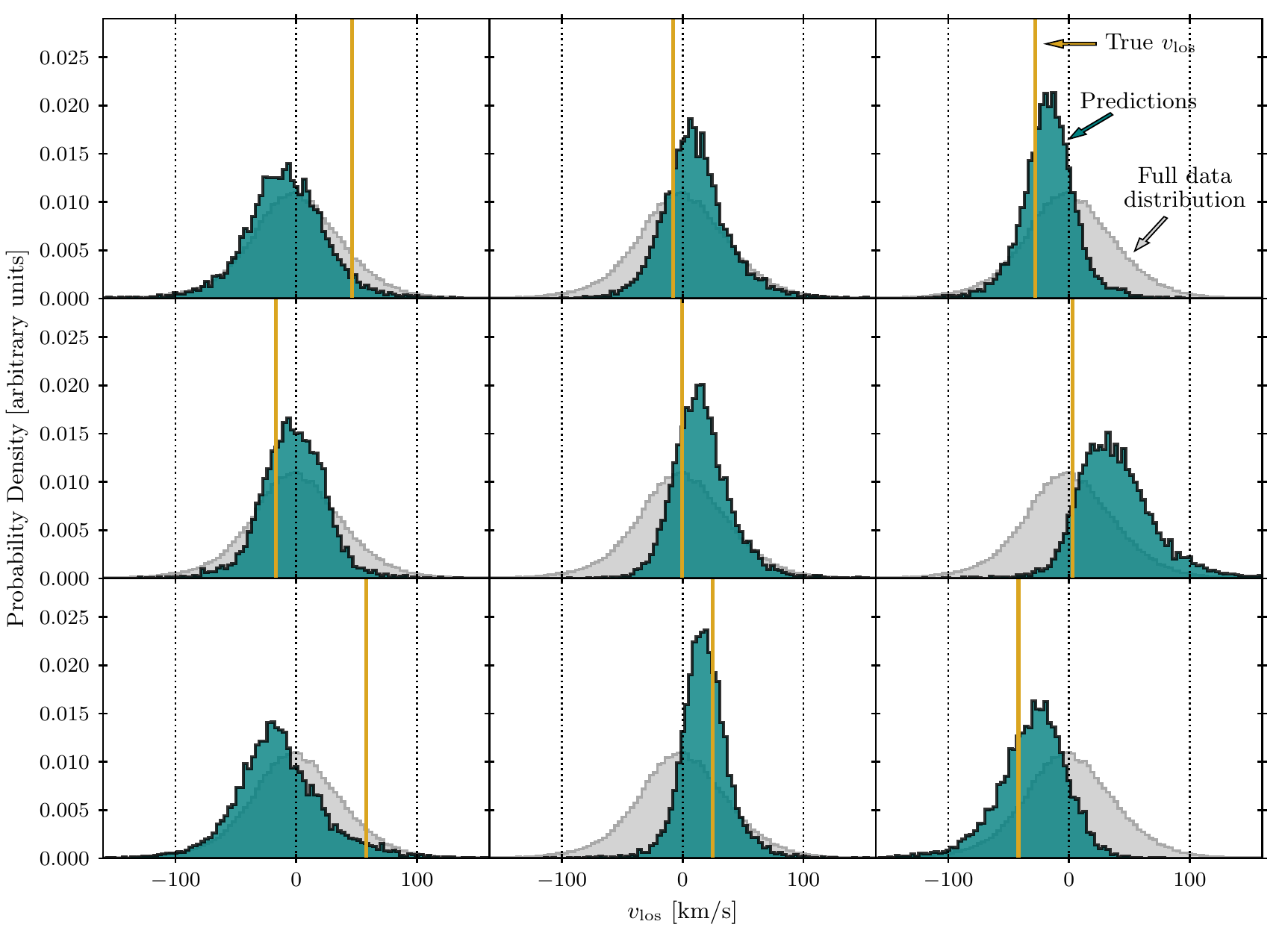}
    \caption{Examples of posterior distributions for $\vlos$ generated by the mock data-trained BNN for nine randomly chosen stars from the test set of the mock \textit{Gaia} EDR3 catalogue of \citet{Rybizki2020}, i.e. mock stars with available radial velocities but not shown to the BNN during training. As labelled in the upper right panel, the yellow vertical line in each case indicates the `true' $\vlos$ for a given star, and the green histogram shows the distribution of 10~000 predictions generated by the BNN. For reference, we also show the $\vlos$ distribution of the full test set (light grey histogram), this is identical in each panel.}
    \label{F:MEDR3PosteriorExamples}
\end{figure*}

\begin{figure*}
    \centering
    \includegraphics{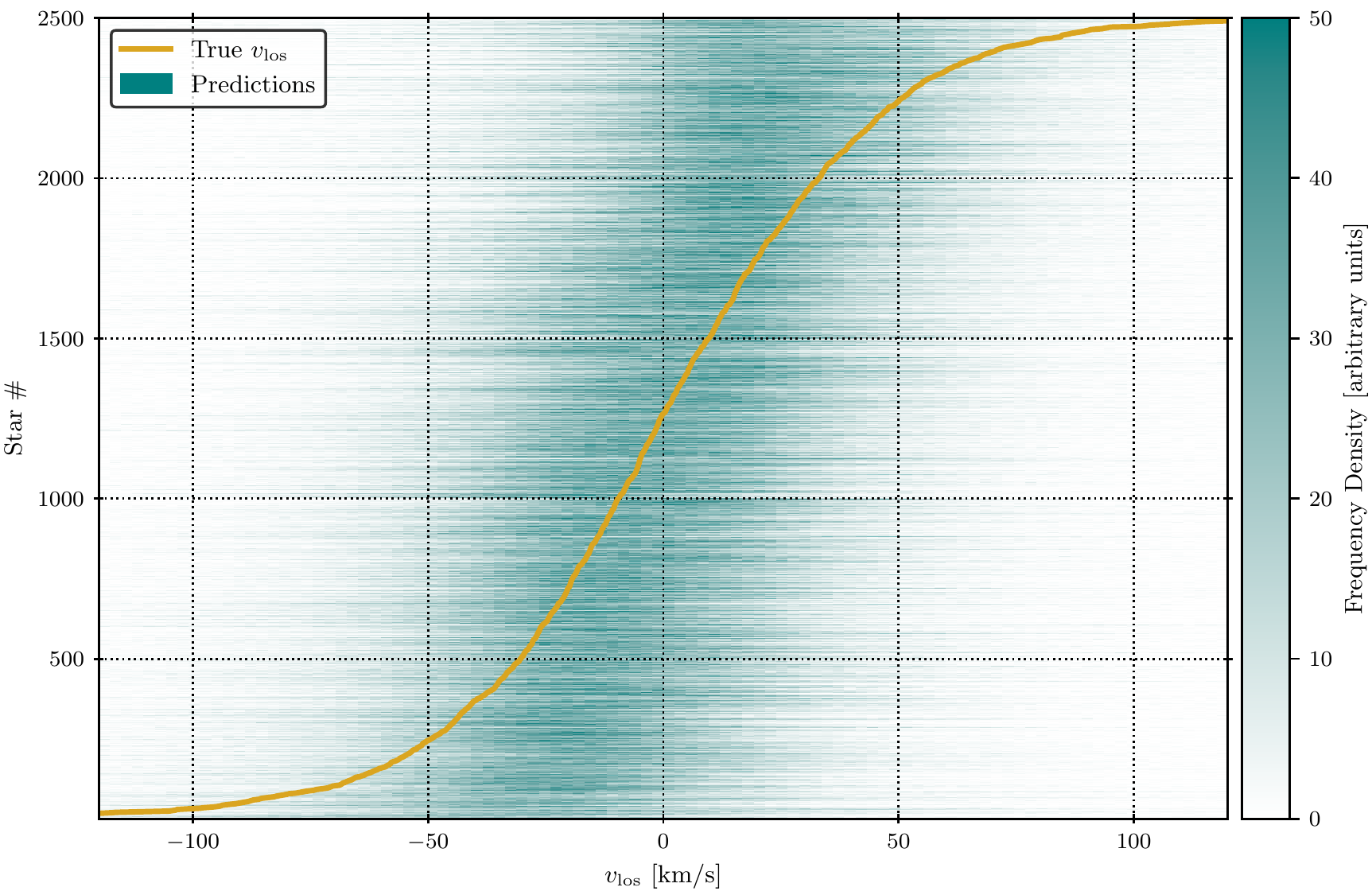}
    \caption{Posteriors generated by the mock data-trained BNN for 2500 randomly chosen stars from the test set of the mock EDR3 catalogue. Each horizontal coloured band represents the posterior of a star $p(\vlos | \bfx, \D_\mathrm{train})$, with darker shading indicating greater probability density. The stars are ordered vertically by their true $\vlos$ values, which are indicated by the yellow line. By eye, it appears that the posteriors are statistically consistent with the true radial velocities. Outliers are pulled toward the top and bottom of the figure and have posteriors centred closer to the global mean, a natural manifestation of `regression toward the mean'.}
    \label{F:MEDR3Posteriors}
\end{figure*}

\begin{figure*}
    \centering
    \includegraphics{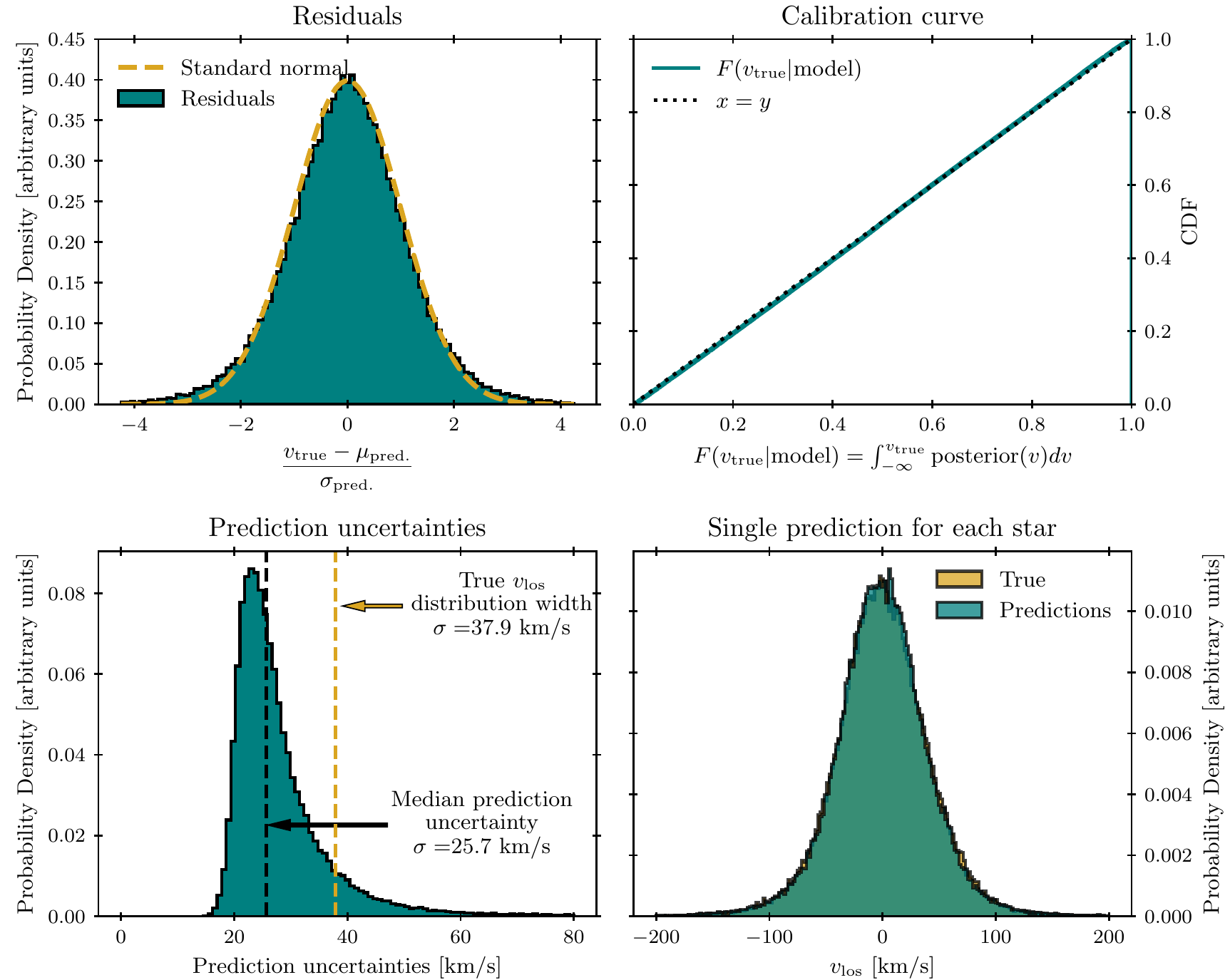}
    \caption{Various measures of accuracy and precision of the mock data-trained BNN predictions against the test set of the mock EDR3 catalogue. \textit{Upper left:} histogram of residuals as defined by Eq.~\ref{E:Residuals}. Also shown for reference: PDF of standard normal distribution (dashed yellow line). \textit{Upper right:} calibration curve (see Eq.~\ref{E:CalibrationCurve} and accompanying discussion). Also shown for reference: $y=x$ (dotted line). \textit{Lower left:} histogram of distribution widths (84\textsuperscript{th} minus 16\textsuperscript{th} percentiles) of predictive posteriors. Vertical black dashed line indicates the median distribution width, and the vertical yellow dashed line indicates the width of the distribution of true radial velocities across the test set. \textit{Lower right:} the true $\vlos$ distribution (yellow histogram) a $\vlos$ distribution generated by randomly drawing a single value from the posterior of each star (teal histogram). Taken together, all of these various findings indicate that the BNN is generating accurate and appropriately confident posteriors for the stars of the test set, with a typical precision of $\sim$25~km/s.}
    \label{F:MEDR3TestStats}
\end{figure*}

\begin{figure*}
    \centering
    \includegraphics{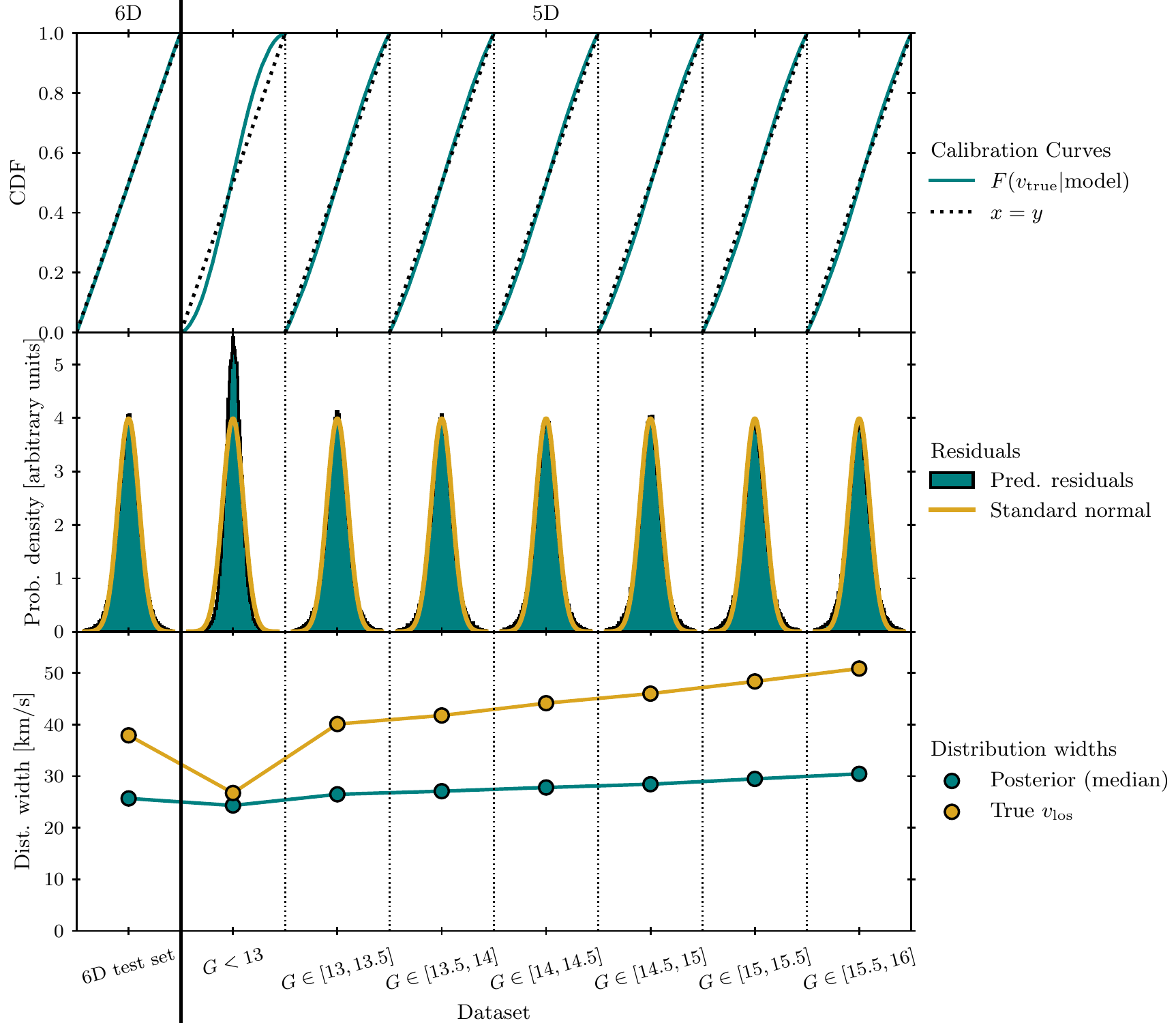}
    \caption{Performance of the mock data-trained BNN on various subsets of the mock stellar population. 8 subsets are shown: the 6D test set (cf. Fig.~\ref{F:MEDR3TestStats}), a bright 5D sample (labelled $G<13$), and six fainter samples (labelled $G \in [13, 13.5], [13.5, 14]$ etc.), as indicated. The 6D and 5D subsets are separated by a solid black partition. \textit{Upper:} calibration curves (see Eq.~\ref{E:CalibrationCurve} and accompanying discussion). Also shown for reference in each case: $y=x$ (dotted lines). \textit{Middle:} histograms of residuals as defined by Eq.~\ref{E:Residuals}. Also shown for reference in each case: PDF of standard normal distribution (dashed yellow lines). \textit{Lower:} The median BNN posterior width for each dataset (teal circles) and the width of the true $\vlos$ distribution (yellow circles). This figure demonstrates the BNN is able to successfully extrapolate its learned model to stars that are fainter than the 6D training set. However, it produces under-confident predictions for the bright 5D sample.}
    \label{F:MEDR3Extrap}
\end{figure*}

\begin{figure*}
    \centering
    \includegraphics{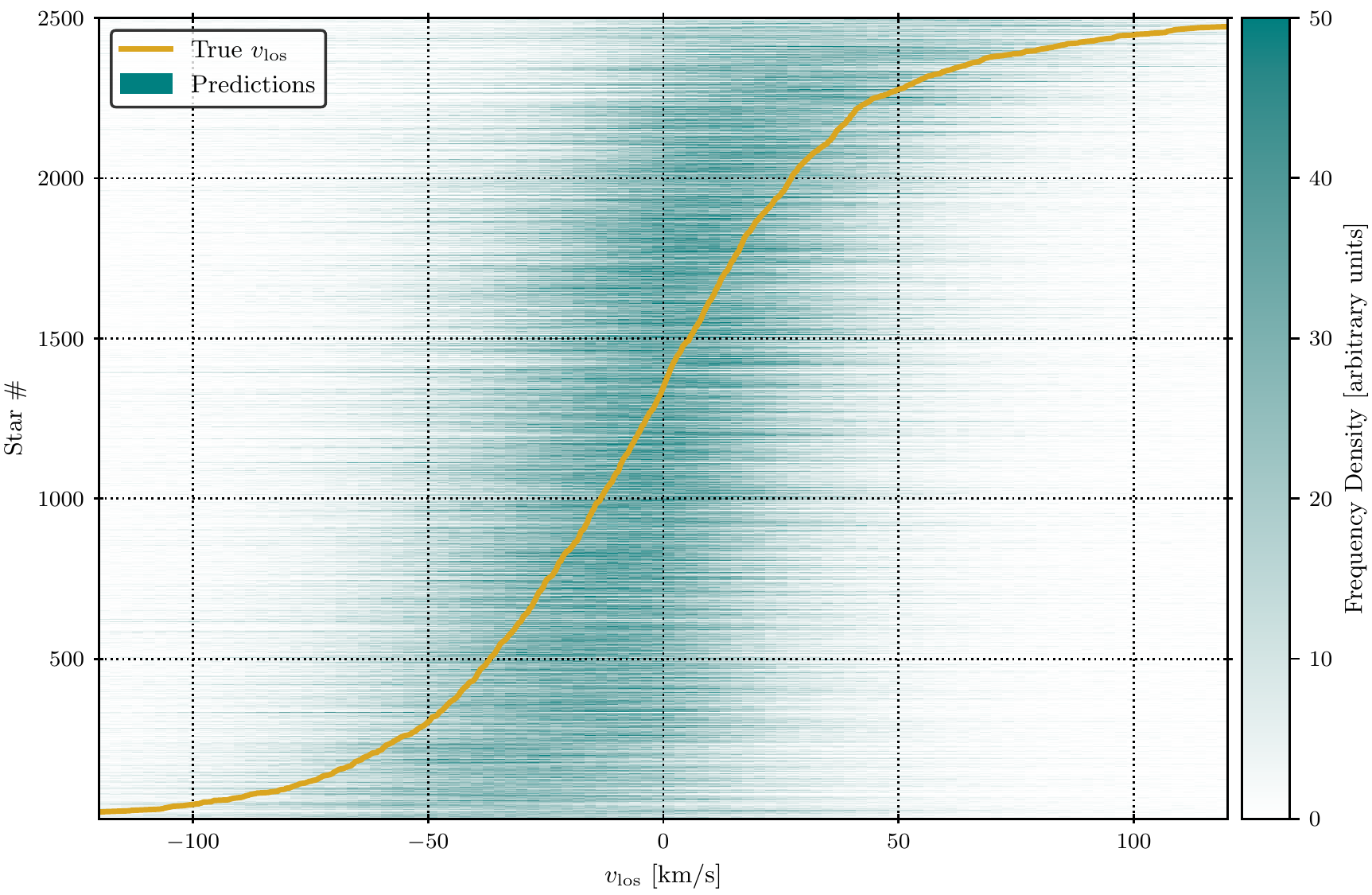}
    \caption{Analogue of Fig.~\ref{F:MEDR3Posteriors}: 2500 BNN-generated posteriors, now for the test set of the real \textit{Gaia} EDR3 stars. As with the mock data, the BNN-generated predictions appear to be statistically consistent with the truths.}
    \label{F:EDR3Posteriors}
\end{figure*}

\begin{figure*}
    \centering
    \includegraphics{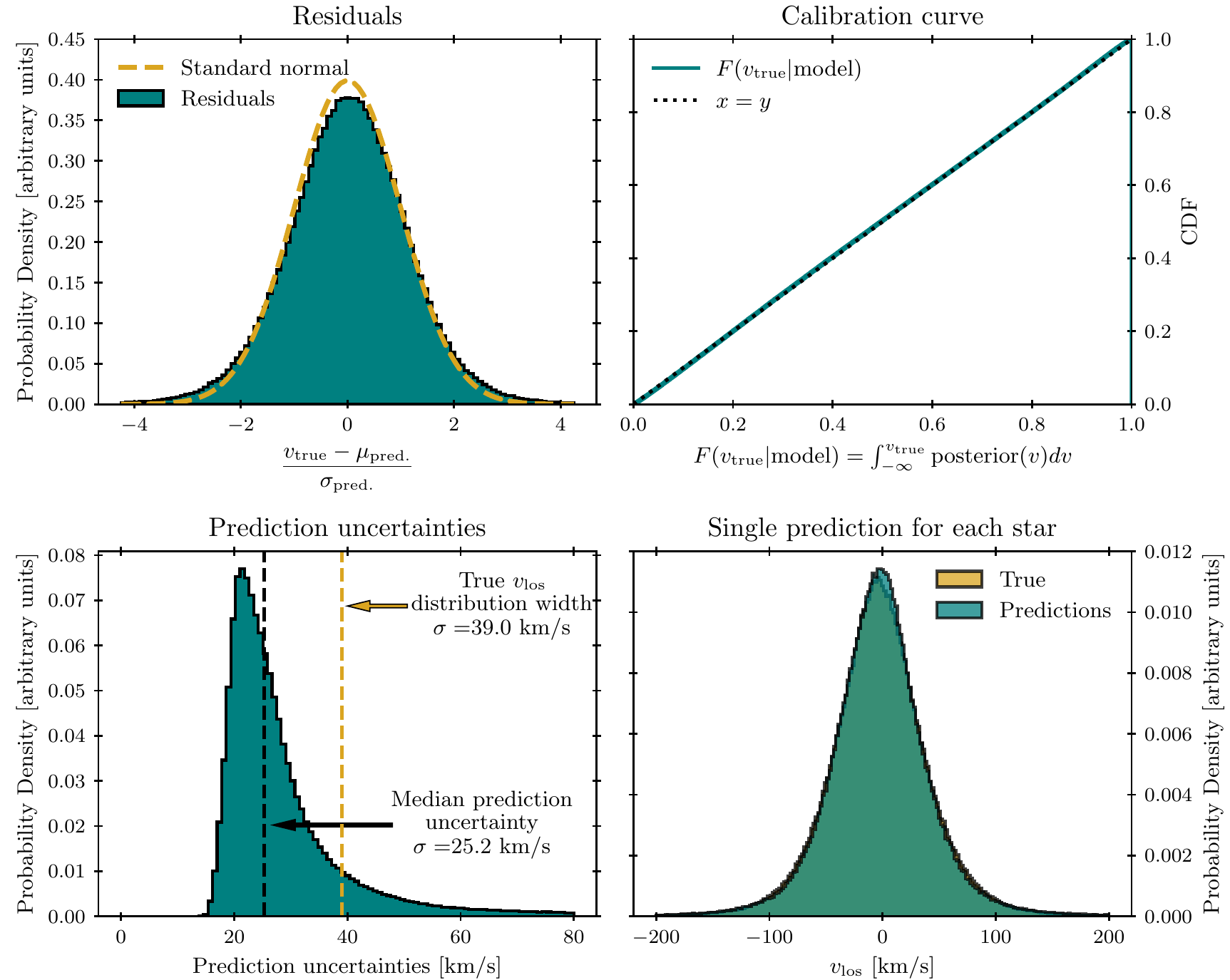}
    \caption{Analogue of Fig.~\ref{F:MEDR3TestStats}: various measures of BNN accuracy and precision, now for the test set of the real \textit{Gaia} EDR3 stars. The performance is almost exactly comparable to the mock data case, with the BNN producing accurate and appropriately confident posteriors for the stars of the test set, with a typical precision of $\sim$25~km/s.}
    \label{F:EDR3TestStats}
\end{figure*}

\begin{figure}
    \centering
    \includegraphics{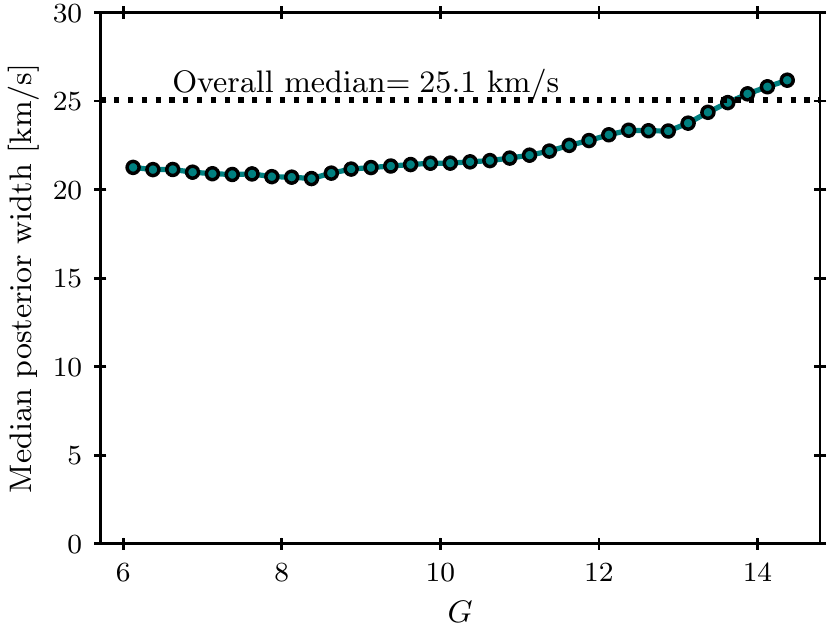}
    \caption{Median $\vlos$ posterior width for 5D EDR3 stars in bins of apparent magnitude $G$ (filled circles). Also shown: overall median across the whole stellar population (dotted line). This figure illustrates the typical uncertainty accompanying our predictions for the DR3 radial velocities is around 25~km/s.}
    \label{F:EDR3PredUncertainties}
\end{figure}

\begin{figure}
    \centering
    \includegraphics{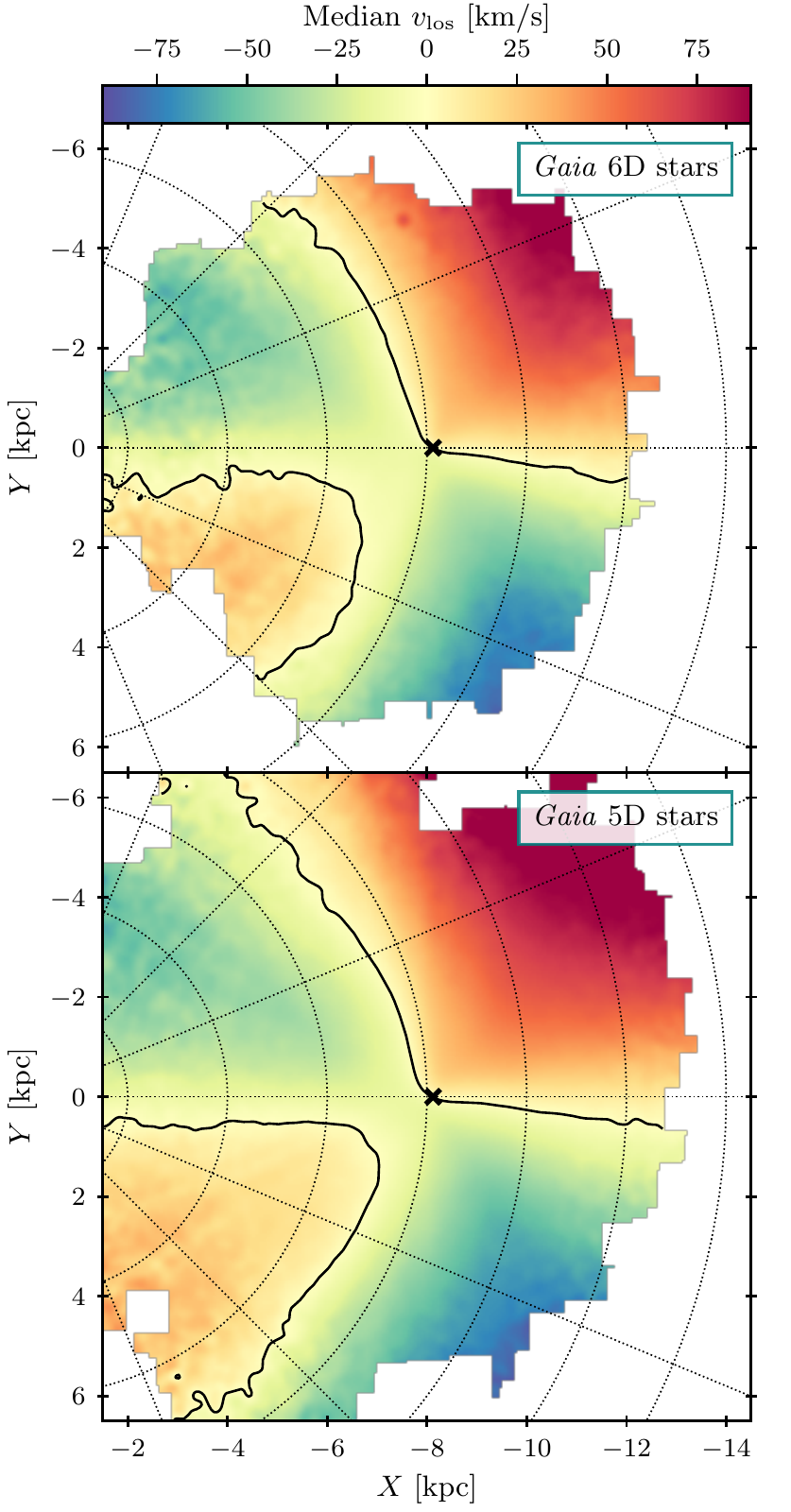}
    \caption{Maps of median $\vlos$ across 2D bins (width 50~pc) in the Galactic plane, \hl{smoothed with a Gaussian filter with a width of 2 pixels}. The axes are oriented and coordinate system defined such that the Galactic centre is to the left of the panels, and the Galactic rotation is clockwise about this. The Sun is indicated by the black cross, \hl{the solid black lines give the contours of zero $\vlos$}, and the dotted grid lines give lines of equal Galactic radius and azimuth. \textit{Top:} map produced from true $\vlos$ of all ($6<G<14.5$) \textit{Gaia}-\texttt{StarHorse} 6D stars. \textit{Bottom:} map produced from all ($6<G<14.5$) \textit{Gaia}-\texttt{StarHorse} 5D stars. In this case, a radial velocity is guessed for each star by drawing a random value from its BNN-generated posterior.}
    \label{F:EDR3RVMap}
\end{figure}

%%%%%%%%%%%%%%%%%%%%%%%%%%%%%%%%%%%%%%%
% MEDR3
\section{Mock Data Validation}
\label{S:MockEDR3}

Before turning to the real \textit{Gaia} data, it is worth exploring the effectiveness of this BNN approach with a mock dataset. There are several benefits to doing so. First, it demonstrates that the technique works, i.e. it is capable of producing accurate, useful predictions for the missing radial velocities. Second, we can use this mock data validation to find the best hyperparameters (e.g. details of the network architecture, number of samples per forward pass $N_s$, etc.) to use when applying to the real data. Finally, we can find the limitations of the technique and learn how far the technique can be extrapolated beyond the 6D subsample.

The mock catalogue we use for this is that of \citet{Rybizki2020}, which has been specifically constructed to resemble \textit{Gaia} EDR3. The catalogue was constructed using the code \textsc{galaxia} \citep{Sharma2011} to sample mock stellar observations from an underlying many-component Galactic model \citep[the Besan\c{c}on model;][]{Robin2003}. The data generation procedure incorporated a state-of-the-art 3D extinction map of the Galaxy, and the data are accompanied by realistic EDR3-like uncertainties.

In the published mock catalogue, all stars have accompanying line-of-sight velocities, not just a bright subset. So, in order to extract a dataset resembling the 6D subsample of the \textit{Gaia} data, we need to emulate the selection function of the \textit{Gaia} RVS. We approximately achieve this with the criteria $G < 13$ and $3550~\mathrm{K} < T_\mathrm{eff.} < 6900~\mathrm{K}$, where $G$ is the apparent $G$-band magnitude and $T_\mathrm{eff.}$ is the effective stellar temperature \citep{Rybizki2018, Katz2019}; we consider the set of stars satisfying these criteria as possessing measured radial velocities, and describe it as our 6D sample. Applying these criteria, the catalogue returns 7\,106\,695 stars, of which we randomly subsample exactly six million. We randomly split this 6D sample into a training set of size 5\,900\,000 and a test set of size 100\,000. 

We construct a BNN as described in Sec.~\ref{S:BNN:Implementation} and train it on the training set using the procedure described in Sec.~\ref{S:BNN:Optimisation}. At the start of every training epoch, we randomly resample the parallaxes, right ascensions, declinations, proper motions and radial velocities of every star from one-dimensional Gaussians with widths equal to the published uncertainties and means equal to the published central values. Note that we are assuming the errors of different quantities are uncorrelated; we adopt a more sophisticated approach for the real \textit{Gaia} data (Sec.~\ref{S:EDR3}). We calculate distance via inverse parallax, then convert the positions to Galactocentric Cartesian coordinates. The calculated quantities then form, for each star, a length-5 input vector $\bfx = (X, Y, Z, \mu_\alpha, \mu_\delta)$ alongside a length-1 output vector $\bfy = (\vlos)$. We rescale the inputs/outputs by subtracting means of $(-8~\mathrm{kpc}, 0, 0, 0, 0)/(0)$ and converting into units of $(0.75~\mathrm{kpc}, 0.75~\mathrm{kpc}, 0.4~\mathrm{kpc}, 15~\mathrm{mas/yr}, 15~\mathrm{mas/yr})/(40~\mathrm{km/s})$.

Post-training, we test the BNN on the test set, which has hitherto remained entirely unseen by the BNN throughout its training. We do this by generating distributions of predictions (i.e. posterior predictive distributions) for the stars of the test set and comparing these predictions with the known truths. Recalling the discussion of Sec.~\ref{S:BNN}, each prediction is generated by randomly resampling the weights and biases $\theta$ of the network from their optimised probability distributions, randomly resampling the input vector $\bfx$ from its uncertainty distribution as described in the previous paragraph, then evaluating the network on the input vector to get a prediction $f_\theta(\bfx)$.

As a preliminary illustration of how this prediction procedure works, Figure~\ref{F:MEDR3PosteriorExamples} plots histograms of 10\,000 predictions for 9 randomly chosen stars from the test set. Properly normalised, these prediction distributions are equivalent to the posterior predictive distributions for the radial velocities of these stars, and we use the terms interchangeably. The shapes of these posteriors demonstrate the utility of our non-parametric optimisation approach: the distributions are often highly skewed. Comparing the prediction distributions with the true $\vlos$ values (also plotted) in these few examples, it is difficult to judge whether the predictions are in good agreement with the truths. For comparison, we also plot the distribution of true radial velocities across the whole test set. For any given star, a first na\"ive guess for its missing radial velocity might be a random value drawn from this full data distribution, but comparing this with the posteriors shows the extent to which can do better with an informed BNN estimate.

To assess the accuracy of the predictions across a broader tranche of the test set, Figure~\ref{F:MEDR3Posteriors} vertically stacks the posteriors for 2500 randomly chosen stars from the test set, ordered by their true $\vlos$ values (as given by the yellow line). \hl{There is a trend visible here}: stars with large positive $\vlos$ have predictions distributions centred around large positive values, and vice versa for stars with large negative $\vlos$. By eye, it appears that the BNN predictions accurately recover the true $\vlos$ values in a statistical sense, i.e. the truths are encompassed by the posteriors. The principle of `regression toward the mean' is evident in the figure, as any outliers with respect to the intrinsic phase space distribution are pulled towards the top and bottom of the figure and have posterior distributions that are centred closer to the global mean.

Figure~\ref{F:MEDR3TestStats} puts this claim of statistical accuracy on a more quantitative footing. In the upper left panel we plot a histogram of residuals across all 100\,000 stars of the test set, where residuals are defined as
\begin{equation}
\label{E:Residuals}
    r = \frac{v_\mathrm{true} - \mu_\mathrm{pred.}}{\sigma_\mathrm{pred.}},
\end{equation}
where $v_\mathrm{true}$ is the true $\vlos$ for a given star, and $\mu_\mathrm{pred.}$ and $\sigma_\mathrm{pred.}$ are the mean and width of its prediction distribution. The width here is defined as half the 84\textsuperscript{th} percentile minus the 16\textsuperscript{th} percentile; we avoid using the standard deviation as many of the posteriors are rather non-Gaussian (cf. Fig~\ref{F:MEDR3PosteriorExamples}). The distribution of $r$ is very close to a standard normal distribution (also plotted for reference), albeit slightly wider: the standard deviation is 1.13. Overall this indicates excellent agreement between the predictions and truths. The slightly increased width of the $r$ distribution can be interpreted in one of two ways: either the posteriors are accurate but under-confident, or the posteriors are sufficiently non-Gaussian that residuals as defined by Eq.~\ref{E:Residuals} should not be expected to exactly obey a normal distribution. To ascertain which of these explanations is correct, we need an alternative metric to assess prediction accuracy. For this, we use quantile-quantile plots, also known as `calibration curves': for each star $i$, we calculate the probability $P_i$ of observing a radial velocity less than the true value, according to our BNN-generated posteriors. To calculate this probability from a set of predictions, we use the same kernel smoothing procedure as described in Sec.~\ref{S:BNN:Implementation} (see Eq.~\ref{E:KDE}), so that
\begin{equation}
\label{E:CalibrationCurve}
    P_i = \int_{-\infty}^{y_{\mathrm{true}, i}} dy p(y | \bfx_i) = \frac{1}{N_s} \sum_j^{N_s}\left[\tanh\left(\frac{y - \hat{y}_{i, j}}{2 h_i}\right) + 1\right],
\end{equation}
where $N_s$ is the number of predictions per star, $h_i$ is the kernel bandwidth for star $i$ (Eq.~\ref{E:Bandwidth}), and $\hat{y}_{i, j}$ is the $j$\textsuperscript{th} prediction for the $i$\textsuperscript{th} star. A calibration curve is then a cumulative density plot of $P_i$ values across a whole dataset. A set of prediction distributions in perfect statistical agreement with the truths would then give a calibration curve falling exactly on the line $y=x$, indicating that e.g. 20\% of stars have true radial velocities in the bottom 20\% of their posteriors, etc. Deviations from $y=x$ warn of deficiencies in the predictive model: over-confident posteriors give N-shaped calibration curves, and under-confident posteriors give S-shaped calibration curves.\footnote{These quantile-quantile plots are defined slightly differently from those of e.g., \citet{Dropulic2021}, who plotted central containment fraction rather than $P_i$ as defined in Eq.~\ref{E:CalibrationCurve}. As a consequence, over/under-confidence leads to different signatures in the plots.} In the upper right panel of Fig.~\ref{F:MEDR3TestStats}, we plot the calibration curve for our test set. The curve falls almost exactly on the $y=x$ line (also plotted), indicating near perfect statistical agreement between predictions and truths.

Beyond the accuracy, it is also interesting to consider the precision of the predictions, which we quantify in the lower left panel of Figure~\ref{F:MEDR3TestStats}. After all, one could achieve the same statistical consistency with the na\"ive guess procedure described above: taking the full distribution of true $\vlos$ values across the dataset as being the predictive distribution for each star. Defining distribution width again as half the 84\textsuperscript{th} percentile minus the 16\textsuperscript{th}, the vertical yellow line in this panel shows the width of this full $\vlos$ distribution across the test set, $37.9~\mathrm{km/s}$. This is to be compared with the histogram, which plots the widths of the prediction distributions for all stars in the test set. These are generally much lower the full data distribution width (median: $25.7~\mathrm{km/s}$), again demonstrating that the BNN predictions do a much better job than the na\"ive guess, by 10-15~km/s.

As a final validation against the test set, in the lower right panel of Figure~\ref{F:MEDR3TestStats} we show a single BNN generation of the overall $\vlos$ distribution, i.e. for each star in the test set we sample a single value from its posterior predictive distribution. Clearly, the generated distribution matches the true distribution very well. Both distributions are close to Gaussian, with the true distribution having mean $-0.36~\mathrm{km/s}$ and standard deviation $42.4~\mathrm{km/s}$, and the generated distribution having mean $-0.52~\mathrm{km/s}$ and standard deviation $45.5~\mathrm{km/s}$.

Given these findings, it thus appears safe to conclude that the BNN has successfully `learned' a model for the radial velocity distribution of the 6D subsample of the mock catalogue, and this model is capable of generating accurate and (comparatively) precise guesses for their radial velocities, given their positions and proper motions. The question now becomes one of how far this model can be extrapolated, i.e. does the BNN produce accurate and precise predictions for the radial velocities of stars lying outside the 6D subsample? The answer to this question will give us a better understanding of exactly which \textit{Gaia} stars we are able to predict radial velocities for.

To address this question, we apply our trained BNN to the 5D stars of the mock catalogue, dividing these stars into multiple subsets. First, there are the bright ($G<13$) stars which lie outside the RVS $T_\mathrm{eff.}$ range of $[3550, 6900]$~K. Then there are the faint ($G>13$) stars, which we split into bins of width 0.5 in apparent magnitude, from $G=13$ to $G=16$. For each subset, we plot in Figure~\ref{F:MEDR3Extrap} several of the same measures of accuracy and precision that we used to test the BNN predictions on the test set. In particular, we plot the calibration curve (top panel), the histogram of residuals as defined in Eq.~\ref{E:Residuals} (middle panel), and the median posterior width and width of the true $\vlos$ distribution, in other words the precisions of the informed and na\"ive guesses respectively (bottom panel).

For the faint star subsets, the predictions continue to perform very well, i.e. the residual histograms are very nearly standard normal distributions, the calibration curves fall very nearly on $y=x$ (with some very slight evidence of under-confidence), and the posterior widths are smaller than the full data distribution widths by 10--20~km/s. This behaviour continues all the way to $G=16$, a full three magnitudes fainter than the magnitude limit of the training set.

The BNN performs slightly less well on the bright ($G<13$) 5D subset. For these stars, the distribution of residuals is rather tighter than a standard normal, and the calibration curve is appreciably S-shaped. This suggests that the BNN is significantly under-confident in its predictions, i.e. the posteriors are broader than they need to be given how closely the distribution centres match the true radial velocities. In a sense, this is undesirable behaviour: an under-confident prediction means that one does not extract the maximum amount of information content available. On the other hand, there are good reasons for this under-confidence. The vast majority of the stars in this subset fall beyond the RVS $T_\mathrm{eff.}$ range at the high end (6900~K); these are young, blue stars and they typically lie much closer to the disc plane than the overall population. As a consequence, they have a significantly different phase space distribution than the overall population. This can be seen in the bottom panel of Figure~\ref{F:MEDR3Extrap}, where the true $\vlos$ spread is much smaller for this subset than for any other subset. One could thus argue that the BNN has been successful in detecting that it is dealing with an out-of-training distribution sample, and is exercising commensurate caution in broadening its prediction distributions. Importantly, this under-confidence is a vastly better outcome than over-confidence, i.e. tight posteriors around incorrect values. By contrast, it appears to be the case that the phase space distribution of the faint samples are sufficiently similar to that of the training set that the predictions retain their confidence, even though the fainter samples stretch to slightly farther distances than the training set (this latter point could be the reason for the slight suggestion of under-confidence in the calibration curves of the faint samples).

%%%%%%%%%%%%%%%%%%%%%%%%%%%%%%%%%%%%%%%
% EDR3
\section{The \textit{Gaia} EDR3 Radial Velocities}
\label{S:EDR3}

Having tested the BNN technique on a realistic mock dataset, we now turn to the real \textit{Gaia} data and generate Bayesian predictions for its missing line-of-sight velocities. 

We download all \textit{Gaia} EDR3 stars in the apparent magnitude range $G\in [6, 14.5]$ which also have accompanying photo-astrometric distance estimates from the \texttt{StarHorse} code \citep{Anders2022}, constituting a total of 23\,213\,405 stars. Of these, 6\,725\,765 are 6D, i.e. they have accompanying radial velocity measurements, leaving a remainder of 16\,487\,640 stars with only five phase space coordinates. We perform a random 90/10 split on the 6D stars to give a training set of 6\,053\,188 stars and a test set of 672\,577 stars. As discussed in Sec.~\ref{S:BNN:Implementation}, we make some further cuts to the training set, removing any stars with large observational uncertainties. In particular, we keep only stars with $\vlos$ errors less than 5~km/s, both proper motion errors less than 0.07~mas/yr, and distance errors less than 0.75~kpc.\footnote{For stellar distances, we use the \texttt{StarHorse} quantity \texttt{dist50} for the central value and $(\mathtt{dist84} - \mathtt{dist16}) / 2$ as a standard deviation representing the distance uncertainty.} Combined, these cuts remove approximately 8.5\% of the training set, leaving 5\,537\,544 stars. Note that we only make these cuts to the training set, not to the test set or the 5D set. For both of these latter sets, we wish to generate predictions for stars regardless of their observational uncertainties.

With this training set in hand, we train a BNN as described in Sec.~\ref{S:BNN:Implementation}. At each training epoch, we resample the distance and $\vlos$ from one-dimensional Gaussians with width equal to the published uncertainty and mean equal to the published central value, and sample the right ascension, declination, and two proper motions of each star from a 4D Gaussian, using the published errors and covariances to fill the $4\times4$ covariance matrix. Given these sampled quantities, we then convert coordinates to construct a length-5 input vector $\bfx = (X, Y, Z, \mu_\alpha, \mu_\delta)$ alongside a length-1 output vector $\bfy = (\vlos)$. We rescale the inputs/outputs by subtracting means of $(-8~\mathrm{kpc}, 0, 0, 0, 0)/(0)$ and converting into units of $(1.5~\mathrm{kpc}, 1.5~\mathrm{kpc}, 0.4~\mathrm{kpc}, 15~\mathrm{mas/yr}, 15~\mathrm{mas/yr})/(40~\mathrm{km/s})$.

Having trained the BNN on the training set, we now turn to the hitherto unseen 6D test set to assess the performance of the BNN, as we did in the previous section with the mock data. Figure~\ref{F:EDR3Posteriors} is the analogue of Figure~\ref{F:MEDR3Posteriors}, now showing 2500 posteriors for the radial velocities of stars in the real \textit{Gaia} test set. Just as in Figure~\ref{F:MEDR3Posteriors}, Figure~\ref{F:EDR3Posteriors} demonstrates that the BNN predictions appear, at least by eye, to be statistically consistent with the truths.

Figure~\ref{F:EDR3TestStats} is then the real data analogue of Fig.~\ref{F:MEDR3TestStats}, quantifying this statistical accuracy. Just as was seen in the mock data case, the residuals (as defined in Eq.~\ref{E:Residuals}) in the upper left panel lie in a distribution that is close to but not quite a standard normal. However, the calibration curve (upper right panel; cf. Eq.~\ref{E:CalibrationCurve}) is near perfect, suggesting that the predictions have the exactly appropriate level of confidence given their accuracy.

The lower left panel of Fig.~\ref{F:EDR3TestStats} then consider the precisions of the BNN predictions for the 6D test set. Again, the numbers are very similar to those obtained with the mock data: the median posterior width is 25.2~km/s, to be compared the width of the true $\vlos$ distribution across the test set, which is 39~km/s. Thus, the BNN gives an improvement in precision of around 15~km/s over the na\"ive first guess.

The lower right panel of of Fig.~\ref{F:EDR3TestStats} plots histograms of the true $\vlos$ distribution across the test set and a generated distribution obtained by taking a drawing a single sample from the posterior of each star. The two distributions agree well, with means of -2.9~km/s and -2.5~km/s respectively and standard deviations of 46.8~km/s and 51.9~km/s respectively.

To summarise, we find that a BNN trained on a portion of the 6D \textit{Gaia} data is able to predict the radial velocities of the remaining portion with excellent accuracy, appropriate confidence, and uncertainties of around 25~km/s. Upon reaching this point with the mock data in Sec.~\ref{S:MockEDR3}, we moved on to testing the ability of the BNN to extrapolate to various 5D samples of the mock catalogue. They were labelled as such because they fell outside the RVS selection criteria, but we nonetheless knew the true radial velocities of these stars and so were able to compare the predictions with ground truths, finding that the BNN was capable of successfully predicting the radial velocities of stars fainter than the RVS magnitude limit and stars outside the RVS $T_\mathrm{eff.}$ limit (although with some under-confidence in the latter case). Of course, for the real \textit{Gaia} data we cannot perform such a test \hl{where the precise truth is known for all stars, although some tests are still viable using complementary ground-based surveys such as RAVE \citep{Steinmetz2006}, APOGEE \citep{Majewski2017}, or LAMOST \citep{Cui2012}}. That being said, the very close similarity of the test set performance with the mock data and the real data (i.e. the resemblance of Figs.~\ref{F:MEDR3Posteriors} \& \ref{F:MEDR3TestStats} to Figs.~\ref{F:EDR3Posteriors} \& \ref{F:EDR3TestStats}) gives us a measure of confidence that the BNN trained on the real data will be able to extrapolate similarly well.

Thus, we use our trained BNN to generate predictions for the 16\,487\,640 stars in our dataset without measured radial velocities. For each star, we draw 250 samples from its posterior. We make available a catalogue containing these posterior samples alongside the \textit{Gaia} \texttt{source\_id} of each star. Alongside this, we also publish a more lightweight catalogue, just listing the \texttt{source\_id} and key percentiles for each posterior: 5\textsuperscript{th}, 16\textsuperscript{th}, 50\textsuperscript{th}, 84\textsuperscript{th}, and 95\textsuperscript{th}. These catalogues are available to download at \url{https://dx.doi.org/10.17639/nott.7216}.

Figs.~\ref{F:EDR3PredUncertainties} and \ref{F:EDR3RVMap} provide some visualisations of these predictions. Fig.~\ref{F:EDR3PredUncertainties} bins the 5D stars in apparent magnitude and plots the median posterior width in each bin. The typical posterior width is around 20~km/s at the bright end ($G\approx 6$) and around 25~km/s at the faint end. This increase in uncertainty as a function of magnitude matches the mock data expectation (cf. Fig.~\ref{F:MEDR3Extrap}). Magnitude is not one of the network inputs, so this dependence on magnitude is not intrinsic to the model; it arises as a result of the different phase-space distributions and larger observational uncertainties of the fainter populations. Also plotted in Fig.~\ref{F:EDR3PredUncertainties} is the median posterior width across the whole 5D dataset: 25.1~km/s (note that the fainter stars dominate the sample, so the overall median matches the medians in the fainter magnitude bins). Meanwhile, Figure~\ref{F:EDR3RVMap} shows maps of median $\vlos$ across 2D bins in the Galactic plane for both the 6D sample (upper panel) and the 5D sample (lower panel), where in the 6D case we have used the true $\vlos$ values and in the 5D case we have generated a single sample from the posterior of each star. The format and scale of this map was purposefully chosen to resemble the map of the \textit{Gaia} radial velocity sample constructed by \citet[][Fig.~7 in that article]{Katz2019}. The BNN predictions recover the spatial distribution of the Galactic velocity distribution very well; the lower panel matches the upper panel and even appears to extrapolate it well to larger distances from the Sun. The correctness of these extrapolations will be tested soon with \textit{Gaia} DR3, but it is at least apparent that by using only informed guesses for the velocities of these stars, we are able to map out the rotational structure of the Galaxy within a few kiloparsecs of the Sun.

%%%%%%%%%%%%%%%%%%%%%%%%%%%%%%%%%%%%%%%
% conclusions
\section{Conclusions}
\label{S:Conclusions}

We have used Bayesian neural networks (BNNs) to predict the radial velocities of stars for which only 5D phase space information (proper motions, sky positions and distances) is available, as is the case for the majority of stars observed by the \textit{Gaia} satellite \citep{Gaia2016mission}. Unlike conventional neural networks, the output of a BNN for a given star is not merely a point prediction for its radial velocity, but rather a series of samples from the posterior probability distribution for its radial velocity.

We first tested this method using the mock catalogue of \citet{Rybizki2020}, designed to resemble the \textit{Gaia} Early Data Release 3 \citep[EDR3;][]{Gaia2021EDR3}. With this mock catalogue, we approximately mimicked the sensitivity range (in apparent $G$-band magnitude and stellar effective temperature) of the \textit{Gaia} Radial Velocity Spectrometer \citep[RVS;][]{Cropper2018, Katz2019} to extract a `6D' sample, i.e. a sample of stars that would have measured radial velocities were this mock catalogue to be real. We further split this sample into `training' and `test' sets, with only the former set being fed to the BNN during the training phase.

After training the BNN on this mock training set, we found it was capable of accurately predicting the radial velocities of the stars in the 6D test set. These predictions are accurate and have the appropriate level of confidence, in the sense that the posteriors are statistically consistent with the truths: $x\%$ of truths are in the bottom $x\%$ of the posteriors, $\forall x$. The typical uncertainty on these predictions (i.e. typical posterior width) is around 25~km/s. These predictions are of course not as informative as actual RV measurements because our predictions have to account for the intrinsic width of the underlying stellar phase-space distribution, but it is nonetheless the case that the BNN posteriors do retain significant predictivity: the posteriors are significantly narrower than the width of the true $\vlos$ distribution (around 40~km/s for the test set).

This good behaviour continues when extrapolating the model, i.e. using the trained BNN to generate predictions for stars which fell outside our RVS-like selection cuts. We tested the BNN on increasingly fainter samples, as far as $G=16$. The BNN predictions remained highly accurate throughout this range, albeit with some small evidence of under-confidence. The typical prediction uncertainties increased from around 25~km/s in the ($G<13$) test set to around 30~km/s at $G=16$. Such an increase in uncertainty is to be expected, as the width of the distribution of true $\vlos$ values increases from around 40~km/s for the former sample to around 50~km/s for the latter.

One qualification to this success was the sample of bright ($G<13$) stars which fell outside the temperature range of the RVS. For these stars, we found significant evidence of the under-confidence in the BNN predictions, i.e. the posteriors were wider than necessary given their proximity to the truths. We ascribe this to the fact that this sample (unlike the faint samples) has a significantly different spatial footprint from that of the training set, such that the BNN exercises more caution in its predictions.

Turning to the real \textit{Gaia} data, we considered all stars with $6 < G < 14.5$ and with accompanying photo-astrometric distance estimates from the \texttt{StarHorse} code \citep{Anders2022}. The 6D subset of this constitutes 6\,725\,765 stars, with the remaining 16\,487\,640 stars not having available radial velocity measurements. As with the mock catalogue, we split the 6D sample into a training set (6\,053\,188 stars, 5\,537\,544 after some quality cuts). and a test set (672\,577 stars). We trained another BNN on this 6D training set, before generating predictions for the 6D test set. As we found with the mock data, the BNN predictions for the 6D test set are in excellent agreement with the truths, demonstrate an appropriate level of confidence, and have a similar degree of precision (around 25~km/s). 

We then used this trained BNN to generate a series of predictions for all 16\,487\,640 stars in our dataset without measured radial velocities, drawing 250 samples from the posterior of each star. The typical prediction uncertainty remains around 25~km/s, with slightly narrower posteriors at brighter magnitudes. We publish these predictions in a catalogue along with the \textit{Gaia} \texttt{source\_id} of each star. We also publish a second, more lightweight catalogue giving only summary statistics for each star. These catalogues are available to download at \url{https://dx.doi.org/10.17639/nott.7216}.

A large proportion of these stars will have measured radial velocities published in the upcoming \textit{Gaia} data release (DR3), due to be released in the week following the submission of this article and publication of the accompanying prediction catalogues. \hl{In an upcoming article, we intend to test our predictions here against the measurements provided in DR3.}

\hl{As with any model, we advise caution when extrapolating. In our case, this would mean relying heavily on our BNN predictions for stars that lie far beyond the DR2/EDR3 6D sample in terms of magnitude, temperature (or colour as a proxy), or spatial location. In such cases, the BNN ought to be able to detect that the stars lie well outside the training distribution and give correspondingly broad prediction distributions, as we saw in our mock data tests. Whether it has been universally successful in doing so with the real data will be tested in our sequel paper.}.

\hl{Alongside our forthcoming article, we also intend to} produce updated catalogues with $\vlos$ predictions for the stars still missing radial velocity measurements in DR3. The catalogues published with this paper thus mainly serve as a way to demonstrate the utility of our technique in a fully transparent manner, thus encouraging the community to utilise it in future: ultimately, the limiting magnitude of the Radial Velocity Spectrometer is brighter than the limiting magnitude of the main \textit{Gaia} astrometric sample, so there will always be gaps to fill in the \textit{Gaia} data. If our blind predictions for the DR3 radial velocities prove consistent with the published values, this will serve as a strong vindication of our Bayesian deep learning approach as a reliable way to do this.

%%%%%%%%%%%%%%%%%%%%%%%%%%%%%%%%%%%%%%%
% acknowledgements
\section*{Acknowledgements}

APN is supported by a Research Leadership Award from the Leverhulme Trust. AW acknowledges support from the Carlsberg Foundation
via a Semper Ardens grant (CF15-0384). This work has made use of data from the European Space Agency (ESA) mission \textit{Gaia} (\url{https://www.cosmos.esa.int/gaia}), processed by the \textit{Gaia} Data Processing and Analysis Consortium (DPAC, \url{https://www.cosmos.esa.int/web/gaia/dpac/consortium}). Funding for the DPAC has been provided by national institutions, in particular the institutions participating in the \textit{Gaia} Multilateral Agreement. We are grateful for access to the University of Nottingham's Augusta HPC service.

%%%%%%%%%%%%%%%%%%%%%%%%%%%%%%%%%%%%%%%
% data availability
\section*{Data Availability}

All data used in this article are publicly available. The mock \textit{Gaia} data used in Sec.~\ref{S:MockEDR3} are hosted at the GAVO Data Center (\url{https://dc.zah.uni-heidelberg.de/}; see \texttt{gedr3mock.main}). The \textit{Gaia} data (and \texttt{StarHorse} distance estimates) used in Sec.~\ref{S:EDR3} are hosted at the AIP \textit{Gaia} server (\url{https://gaia.aip.de/}). Our prediction catalogues are hosted at the Nottingham Research Data Management Repository (\url{https://dx.doi.org/10.17639/nott.7216}).

%%%%%%%%%%%%%%%%%%%%%%%%%%%%%%%%%%%%%%%
% bibliography
\bibliographystyle{mnras}
\bibliography{library}

%%%%%%%%%%%%%%%%%%%%%%%%%%%%%%%%%%%%%%%
% doc end --- don't change these lines
\bsp
\label{lastpage}
\end{document}